\documentclass[aps,prl,twocolumn,superscriptaddress,showpacs,showkeys,10pt,longbibliography]{revtex4-1}
\usepackage{graphicx}
\usepackage{textcomp}
\usepackage{color}
\usepackage{dcolumn}
\usepackage{multirow}
\usepackage{amsmath}
\usepackage{amssymb}
\newcommand{\refeq}[1]{Eq.~(\ref{#1})}
\newcommand{\reffig}[1]{Fig.~\ref{#1}}

\usepackage[colorlinks]{hyperref}
\hypersetup{
	plainpages=true,
	breaklinks=true,
	hypertexnames=false,
	pageanchor=true,
	colorlinks=true,
	linkcolor={blue},
	citecolor={blue},
	urlcolor={blue},
	anchorcolor={black}
}

\begin{document}

\title{Electromagnetically induced transparency in circuit quantum electrodynamics with nested polariton states}
\author{Junling Long}\email{Junling.Long@nist.gov}
\affiliation{National Institute of Standards and Technology, Boulder, Colorado 80305, USA}
\affiliation{Department of Physics, University of Colorado, Boulder, Colorado 80309, USA}
\author{H. S. Ku}
\affiliation{National Institute of Standards and Technology, Boulder, Colorado 80305, USA}
\author{Xian Wu}
\affiliation{National Institute of Standards and Technology, Boulder, Colorado 80305, USA}
\author{Xiu Gu}
\affiliation{Institute of Microelectronics, Tsinghua University, Beijing 100084, China}
\author{Russell E. Lake}
\affiliation{National Institute of Standards and Technology, Boulder, Colorado 80305, USA}
\author{Mustafa Bal}
\affiliation{National Institute of Standards and Technology, Boulder, Colorado 80305, USA}
\author{Yu-xi Liu}
\affiliation{Institute of Microelectronics, Tsinghua University, Beijing 100084, China}
\affiliation{Tsinghua National Laboratory for Information Science and Technology (TNList), Beijing 100084, China}
\author{David P. Pappas}
\affiliation{National Institute of Standards and Technology, Boulder, Colorado 80305, USA}

\date{\today}

\begin{abstract}
Quantum networks will enable extraordinary capabilities for
communicating and processing quantum information.  These networks
require a reliable means of storage, retrieval, and manipulation of
quantum states at the network nodes.  A node receives one or more
coherent inputs and sends a conditional output to the next cascaded
node in the network through a quantum channel. Here, we demonstrate
this basic functionality by using the quantum interference mechanism
of electromagnetically induced transparency in a transmon qubit
coupled to a superconducting resonator. First, we apply a microwave bias, i.e., drive, to the the qubit--cavity system to prepare a $\Lambda$-type
three-level system of polariton states. Second, we input two interchangeable microwave signals, i.e., a
probe tone and a control tone, and observe that transmission of the probe tone is conditional upon the
presence of the control tone that switches the state of the device with
up to 99.73 \% transmission extinction. Importantly, our EIT scheme uses
all dipole allowed transitions. We infer high dark state
preparation fidelities of $>$ 99.39 \% and negative group velocities of
up to $-0.52\pm0.09$~km/s based on our data.
\end{abstract}


\maketitle

Controllable interaction between electromagnetic quanta and discrete
levels in a quantum system, i.e., light matter interaction, is the key to quantum information storage
and processing in a quantum
network~\cite{cirac1997quantum,kimble2008quantum}. Consider a
three-level atomic system driven by two coherent electromagnetic
waves. The destructive interference between the two excitation
pathways creates a transparency window for one of the drive fields and
switches the system into a ``dark state.'' This phenomenon is called
electromagnetically induced transparency
(EIT)~\cite{fleischhauer2005electromagnetically}. Recently, EIT has
been harnessed for implementing different building blocks of a quantum
network, such as all-optical switches and
transistors~\cite{baur2014single,shomroni2014all,bajcsy2009efficient,chen2013all,souza2013coherent},
quantum storage
devices~\cite{liu2001observation,phillips2001storage,van2003atomic,kuzmich2003generation,chaneliere2005storage},
and conditional phase
shifters~\cite{tiarks2016optical,beck2016large,shahmoon2011strongly,ottaviani2003polarization,liu2016large}. Despite
this remarkable success, utilizing EIT and related effects at the
single-photon and single-atom level with highly scalable devices is a
formidable challenge that prevents realization of a practical quantum
network~\cite{ma2017optical}. A promising solution is to extend these
techniques to the microwave domain using superconducting quantum
circuits that are both scalable and enable deterministic placement of
long-lived artificial atoms for the network nodes~\cite{kirchmair2012observation,blais2004cavity,schuster2006resolving,koch2007charge}.

To this end, three-level superconducting artificial atoms have been
used to demonstrate coherent population trapping
(CPT)~\cite{kelly2010direct} and Autler-Townes splittings
(ATS)~\cite{baur2009measurement,sillanpaa2009autler,abdumalikov2010electromagnetically,novikov2013autler,hoi2013giant,suri2013observation,cho2014quantum}. However,
conclusive evidence of EIT in these simple systems eluded researchers
as it is difficult to find a superconducting quantum system with
metastable states and lifetimes that satisfy its stringent
requirements~\cite{anisimov2011objectively,sun2014electromagnetically,peng2014and,liu2016method}. Recently,
progress has been made in a circuit quantum electrodynamics (QED)
system that exploits qubit coupling to a single-mode
cavity~\cite{novikov-natphys-2016}.  In that experiment, one leg of
the $\Lambda$-type system is dipole forbidden, requiring that it be
driven with a two-photon transition.  The small photon scattering cross section of this
two-photon transition hinders applications such as single atom quantum
memory~\cite{specht2011single}, all-optical switching and routing of a
single photon gated by another single photon~\cite{shomroni2014all},
single photon-photon cross phase modulation~\cite{hoi2013giant}, and
vacuum induced transparency~\cite{tanji2011vacuum}.  On the other
hand, high scattering cross sections have been observed in a dipole
allowed transition of an artifacial atom coupled to one dimensional
waveguide~\cite{astafiev2010resonance}.  Thus, implementing a
$\Lambda$-type system with all dipole allowed transitions in a circuit
QED system is highly desirable for building a quantum network with
microwave photons.

In this Letter, we report the first observation of EIT using all dipole allowed transitions in a $\Lambda$-type system implemented with superconducting quantum circuit. Our scheme is based on a theoretical proposal~\cite{gu2016polariton} that utilizes polariton states generated with a rf biased two-level system coupled to a resonator. Here, we realized the polariton states in a transmon--cavity system and achieved a metastable state with a long lifetime. Moreover, we were able to tune the polariton states to establish a $\Lambda$-type system that can be driven with control- and probe-fields through dipole allowed transitions. Note that due to the transmission geometry of our cavity where nominally the signal is transmitted on resonance, the observed experimental signal is actually electromagnetically induced absorption (EIA). However, our EIA and conventional EIT have identical underlying physics of quantum interference. Conventional EIT spectra can be observed if a hanger resonator geometry is used. We retain the nomenclature of quantum optics and use the term ``EIT'' for the rest of the Letter. From our EIT data, a large transmission extinction (99.73 \%) of the probe field is observed and high dark state preparation fidelity ($>$ 99.39~\%) is inferred. To our best knowledge, the EIT transmission extinction of 99.73 \% is the highest one that has been observed to date in the circuit QED system. Our EIT scheme opens up new possibilities for realizing scalable devices that utilize single-photons and single-atoms for constructing EIT as a building block of a quantum network in the microwave domain.

\begin{figure}
 \includegraphics{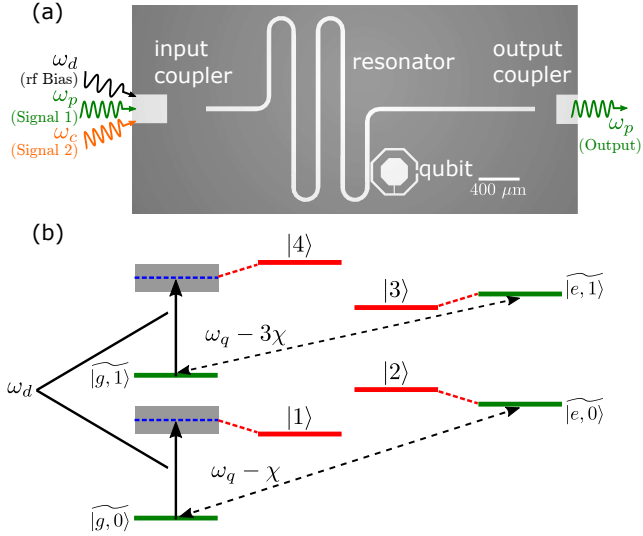}
 \caption{(a) Optical micrograph (with false color) of the device including a capacitively coupled $\lambda/2$ microstrip resonator and a concentric transmon qubit. $\omega_d$, $\omega_p$, and $\omega_c$ are the frequencies of the polariton drive field, the transmission spectrum (or EIT) probe field, and the EIT control field, respectively. (b) Generation of polariton states (red solid lines) in the nesting regime from the Jaynes-Cummings ladder (green solid lines). Black dashed double-headed arrows indicate the photon-number-dependent qubit transitions, black solid arrows shows the polariton drive, and the shaded regions denote the nesting regime. $\omega_q$ and $\chi$ represent the bare qubit frequency and the effective dispersive shift, respectively. Microwave fields are applied through the input coupler in (a). \label{Schematics}}
\end{figure}

Our experiment is performed on the device that consists of a concentric transmon capacitively coupled to a $\lambda/2$ microstrip resonator with a coupling strength $g/2\pi=74$~MHz, as shown in \reffig{Schematics}(a).  The transmon comprises a single Al/AlO$_x$/Al Josephson junction shunted by a capacitor consisting of a superconducting island and a surrounding ring. The Josephson junction is fabricated with an overlap technique \cite{Wu2017}. The transmon has a resonance frequency $\omega_q/2\pi=5.648$~GHz between its lowest two levels and an anharmonicity $\alpha/2\pi=262.5$~MHz. The coherence times are measured to be $T_{1} = 35~\mu$s and $T_{2}^{*}=22.5~\mu$s. The fundamental mode of the resonator is at $\omega_r/2\pi=6.485$~GHz with a internal quality
factor $Q_{i}=(1.2\pm0.2)\times10^{6}$ and a loaded quality factor $Q=7,900$ dominated by the strong coupling to the microwave feedline at the output port.

The transmon-cavity system is well in the dispersive regime with a dispersive shift $\chi/2\pi=1.54$~MHz. The eigenlevels are described by the dispersive Jaynes-Cummings ladder as shown in \reffig{Schematics}(b). The resonance frequencies are $\omega_q-\chi$ for the $\widetilde{\left|g,0\right\rangle}\leftrightarrow\widetilde{\left|e,0\right\rangle}$ transition, and $\omega_q-3\chi$ for the $\widetilde{\left|g,1\right\rangle}\leftrightarrow\widetilde{\left|e,1\right\rangle}$ transition, where $\widetilde{\left|g,n\right\rangle}$ ($\widetilde{\left|e,n\right\rangle}$) denotes the qubit ground (excited) state with $n$ photons in the resonator. The tilde indicates that these levels are singly-dressed states, i.e., they are transmon states slightly dressed with resonator photons.

The polariton states are generated by injecting a strong microwave drive field through the input coupler to doubly dress the Jaynes-Cummings states. In particular, if the drive frequency $\omega_d$ is in the so-called nesting regime, $\omega_q-3\chi<\omega_d<\omega_q-\chi$, the resulting eigenstates $\left|2\right\rangle$ and $\left|3\right\rangle$ will be nested in between the eigenstates $\left|1\right\rangle$ and $\left|4\right\rangle$~\cite{koshino2013implementation,inomata2014microwave,inomata2016single}. 

\begin{figure}
 \includegraphics{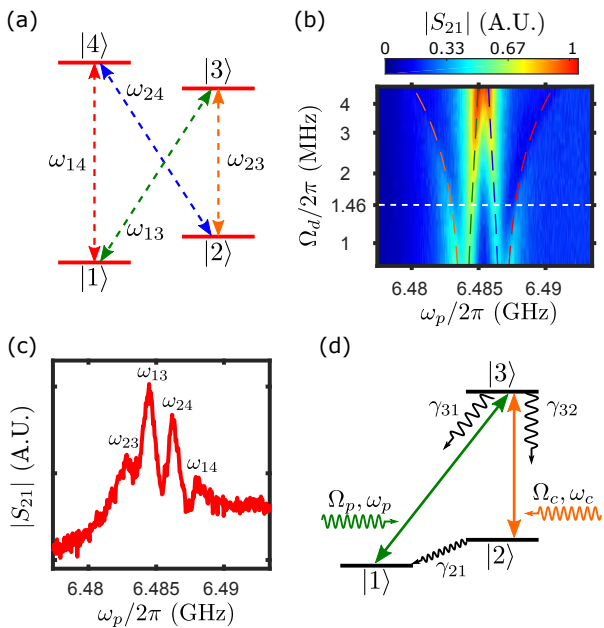}
 \caption{Transmission spectrum of polariton states in the nesting regime. (a) The four transitions (dashed double-headed arrows) between the polariton states. (b) Transmission spectrum of polariton states in arbitrary units shows the four different transitions in (a). $\Omega_d$ and $\omega_p$ are the Rabi strength of polariton drive and the probe frequency, respectively. Dashed curves denote predicted transmission peaks using the AC Stark shift model. (c) A line cut on (b) at $\Omega_d/2\pi=1.46$~MHz. (d) The lowest three levels, $\left|1\right\rangle$, $\left|2\right\rangle$, and $\left|3\right\rangle$, form the $\Lambda$-type transition for implementing EIT. $\Omega_p$ ($\Omega_c$) and $\omega_p$ ($\omega_c$) are the Rabi strength and frequency of the probe (control) field in an EIT experiment, respectively.
 \label{spec}}
\end{figure}

We use the set of polariton states $\left|1\right\rangle$, $\left|2\right\rangle$, and $\left|3\right\rangle$ to form a $\Lambda$-type system [\reffig{spec}(d)]. In the driven two-level-system model, these polariton states can be approximated as
\begin{equation}
\begin{split}
 &\left|1\right\rangle=-\sin\left(\frac{\theta_0}{2}\right)\widetilde{\left|e,0\right\rangle}+\cos\left(\frac{\theta_0}{2}\right)\widetilde{\left|g,0\right\rangle},\\
 &\left|2\right\rangle=\cos\left(\frac{\theta_0}{2}\right)\widetilde{\left|e,0\right\rangle}+\sin\left(\frac{\theta_0}{2}\right)\widetilde{\left|g,0\right\rangle},\\
 &\left|3\right\rangle=-\sin\left(\frac{\theta_1}{2}\right)\widetilde{\left|g,1\right\rangle}+\cos\left(\frac{\theta_1}{2}\right)\widetilde{\left|e,1\right\rangle},\\
 &\left|4\right\rangle=\cos\left(\frac{\theta_1}{2}\right)\widetilde{\left|g,1\right\rangle}+\sin\left(\frac{\theta_1}{2}\right)\widetilde{\left|e,1\right\rangle},
\end{split}
\label{Polarition states}
\end{equation}
where the mixing angles $\theta_0$ and $\theta_1$ are given by, $\tan(\theta_0)=\Omega_d/[(\omega_q-\chi)-\omega_d]$ and $\tan(\theta_1)=\Omega_d/[\omega_d-(\omega_q-3\chi)]$~\cite{gu2016polariton}.

\refeq{Polarition states} shows that the $\left|1\right\rangle\leftrightarrow\left|3\right\rangle$ and $\left|2\right\rangle\leftrightarrow\left|3\right\rangle$ transitions are mainly cavity-like transitions, while $\left|1\right\rangle\leftrightarrow\left|2\right\rangle$ is a qubit-like transition. These properties can be revealed by calculating the decay rate $\gamma_{ij}$ of $\left|i\right\rangle\rightarrow\left|j\right\rangle$ transition, which can be approximated as, $\gamma_{31}=\gamma_c\sin^2\left[(\theta_0+\theta_1)/2\right]$, $\gamma_{32}=\gamma_c\cos^2\left[(\theta_0+\theta_1)/2\right]$, and $\gamma_{21}=\gamma_q\cos^4\left(\theta_0/2\right)$, where $\gamma_c$ is the cavity decay rate and $\gamma_q$ is the qubit decay rate~\cite{gu2016polariton}. Thus, the decay rate of $\left|3\right\rangle\rightarrow\left|1\right\rangle$ transition ($\gamma_{31}$) can be tuned to be comparable with the decay rate of $\left|3\right\rangle\rightarrow\left|2\right\rangle$ transition ($\gamma_{32}$), while extending the metastable state lifetime ($1/\gamma_{21}$) even beyond the qubit lifetime. These two effects are key to achieve EIT in our superconducting circuit system.

We measure the transition frequencies between the polariton states by performing two-tone spectroscopy with a polariton drive and a
weak probe field. The drive frequency and the probe power are fixed at $\omega_d/2\pi=5.6466$~GHz and P$_p=-163.15$~dBm respectively, while scanning the drive strength and the probe frequency. The probe transmission, defined as the ratio of the probe output complex amplitude to the input complex amplitude $S_{21}\equiv V_\mathrm{out}/V_\mathrm{in}=|S_{21}|e^{i\phi}$, was measured by a vector network analyzer (VNA). Our definition of $S_{21}$ includes all round-trip amplification and attenuation, where $\phi$ has been corrected for electric delay. As shown in \reffig{spec}(b), there are four transmission peaks near the resonator frequency. The four peaks correspond to, from low to high frequencies, $\omega_{23}$, $\omega_{13}$, $\omega_{24}$, and $\omega_{14}$ respectively [\reffig{spec}(c)], where $\omega_{ij}$ denotes the energy difference between the polariton states $\left|i\right\rangle$ and $\left|j\right\rangle$. The spacing between the first and second (first and third) transmission peaks, which corresponds to the splitting between levels $\left|1\right\rangle$ and $\left|2\right\rangle$ ($\left|3\right\rangle$ and $\left|4\right\rangle$), widens as the drive strength increases. This is consistent with the expected AC Stark shift drawn as the black dashed curves in~\reffig{spec}(b). Another crucial feature of the spectrum is that, as the polariton drive strength increases, the height of the $\omega_{23}$ and $\omega_{14}$ peaks decreases, while the height of $\omega_{13}$ and $\omega_{24}$ peaks increases. This behavior agrees with the change of the transition probabilities between polariton states predicted in reference~\cite{gu2016polariton}.

\begin{figure}
 \includegraphics{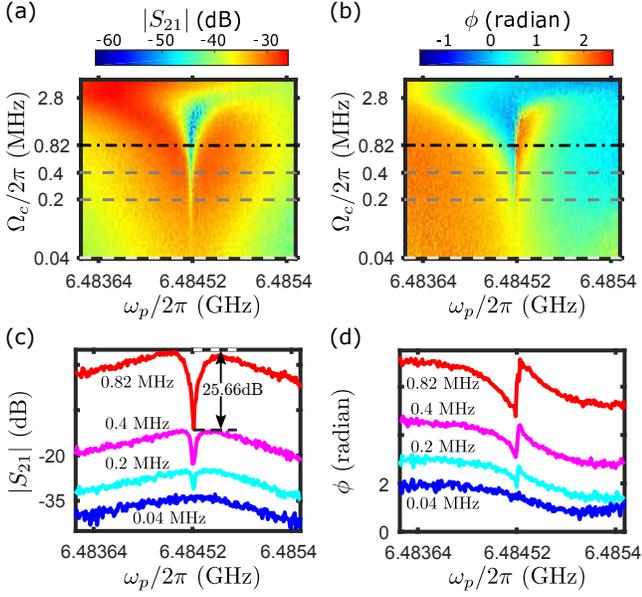}
 \caption{Transmission magnitude (a) and phase (b) of EIT with varying control field strength. (c) and (d) are line cuts on (a) and (b) with $\Omega_c/2\pi=0.04$~MHz, $0.2$~MHz, $0.4$~MHz, and $0.82$~MHz, respectively. The black dash-dotted lines in (a) and (b) denote the boundary of EIT set by $\Omega_c/2\pi=\gamma_c/2\pi=0.82$~MHz. Note that traces are offset vertically from the $\Omega_c/2\pi=0.04$~MHz case for (c)\&(d), and $S_{21}$ data include all round-trip amplification and attenuation.\label{EIT}}
\end{figure}

In this system, EIT is demonstrated by a suppression of transmission for a weak probe field on resonance with one leg of a $\Lambda$-system, while a control field addressing the other leg [\reffig{spec} (d)]. The $\Lambda$-system is established by a polariton drive field with frequency $\omega_d/2\pi=5.6466$~GHz and strength $\Omega_d/2\pi=1.46$~MHz. The resultant $\Lambda$ levels have $\gamma_{31}/2\pi=0.35$~MHz and $\gamma_{32}/2\pi=0.47$~MHz, which are much larger than $\gamma_{21}/2\pi=2.74$~kHz. The control field frequency $\omega_c/2\pi=\omega_{23}/2\pi=6.4828$~GHz and the probe strength $\Omega_p/2\pi=62$~kHz are fixed, while we scan the control field strength $\Omega_c$ and the probe frequency $\omega_p$. The probe transmission ($S_{21}$) measured by the VNA is shown in \reffig{EIT}(a)\&(b). With our parameters, the theoretical condition of EIT is given by $\Omega_c/2\pi<\gamma_c/2\pi=0.82$~MHz [black dash-dotted line in \reffig{EIT}(a)] \cite{gu2016polariton}. Under this condition, we observe a transmission suppression window around $\omega_p=\omega_{13}$ with the largest suppression 25.66~dB [\reffig{EIT}(c)\&(d)], which means about 99.73\% of power of the original transmitted probe field is suppressed. However, as the control field strength exceeds the EIT boundary, the transmission for $\omega_p>\omega_{13}$ in \reffig{EIT}(a) is becoming smaller and completely disappears above $\Omega_c/2\pi=2.8$~MHz instead of changing to an ATS lineshape. This behavior is most likely due to excess cavity population, above a single photon, due to the strong control field.

Quantum interferences in a driven $\Lambda$-system create a dark state, which is transparent to the probe field. The fidelity of the dark state preparation is an important metric for a EIT-based quantum memory \cite{ma2017optical}. With our experimental parameters, we inferred the dark state fidelity defined as \cite{Li2011ATS}
\begin{equation}
\begin{split}
 &F_{\left|D\right\rangle}=\sqrt{\left\langle D\right| \rho \left|D\right\rangle},\\ 
 &=\sqrt{\frac{1}{2}\left[\cos{2\Theta}(\rho_{11}-\rho_{22})-\sin{2\Theta}(\rho_{21}+\rho_{12})+(1-\rho_{11})\right]},
 \end{split}
\end{equation}
where the dark state $\left|D\right\rangle=\cos{\Theta}\left|1\right\rangle-\sin{\Theta}\left|2\right\rangle$ and the mixing angle $\Theta=\mathrm{tan}^{-1}(\Omega_p/\Omega_c)$. The density matrix $\rho$ is calculated by numerically solving a Lindblad master equation of a driven $\Lambda$-system, including decay rates $\gamma_{ij}$ \cite{JOHANSSON2013}. At the EIT boundary ($\Omega_p/2\pi=0.82$~MHz, $\Omega_c/2\pi=0$~MHz), the dark state fidelity is calculated to be 99.39~\%. Note that we switched the role of the probe and the control fields to simulate the fidelity when the dark state is essentially the polariton $\left|2\right\rangle$ state and the main infidelity source is its decay rate $\gamma_{21}$.

To confirm that the suppression of transmission is due to EIT, as opposed to ATS, Akaike’s information criterion (AIC) based testing was performed. The AIC-based testing calculates the weight of each fitting model based on the goodness of the fitting with the constraint that sum of the weights is unity~\cite{anisimov2011objectively}. Originally, the AIC-based testing was proposed to fit the susceptibility, $\chi_s$~\cite{anisimov2011objectively}. To use this criterion, we derive the relationship between the measured $S_{21}$ and a generic susceptibility $\chi_s$ as~\cite{jackson_classical_1999}
\begin{equation}
 \ln(S_{21})=\ln(|S_{21}|)+i\phi=i\frac{\omega_pL}{c}\left(1+\frac{1}{2}\chi_s\right)-\alpha_0+i\phi_0,\label{s2chi}
\end{equation}
where $L$ is the effective distance the microwave travels through the chip, $c$ is the speed of light, $\alpha_0$ is the attenuation of the cables, and $\phi_0$ is a frequency-independent initial phase offset. For EIT, the susceptibility takes the form, the difference between two Lorentzians~\cite{anisimov2011objectively},
\begin{equation}
\chi_{s}^{\mathrm{EIT}}=\frac{A_{+}}{(\omega_p-\omega_{+})-i(\Gamma_{+}/2)}-\frac{A_{-}}{(\omega_p-\omega_{-})-i(\Gamma_{-}/2)},
\label{EITchi}
\end{equation}
and for ATS, it takes the form of the sum,
\begin{equation}
\chi_{s}^{\mathrm{ATS}}=\frac{A_{1}}{(\omega_p-\omega_{1})-i(\Gamma_{1}/2)}+\frac{A_{2}}{(\omega_p-\omega_{2})-i(\Gamma_{2}/2)},
\label{ATSchi}
\end{equation}
where $\omega_j$, $A_j$, and $\Gamma_j$ are the center frequency, magnitude, and width of the $j$th Lorentzian, respectively. In comparison to reference~\cite{anisimov2011objectively}, the different negative signs in front of the $i(\Gamma_{j}/2)$ terms in \refeq{EITchi} and (\ref{ATSchi}), are due to the transmission geometry of the circuit. The model functions for EIT or ATS are then obtained by substituting either $\chi_{s}^{\mathrm{EIT}}$ or $\chi_{s}^{\mathrm{ATS}}$ for the $\chi_s$ in \refeq{s2chi}.

\begin{figure}
 \includegraphics{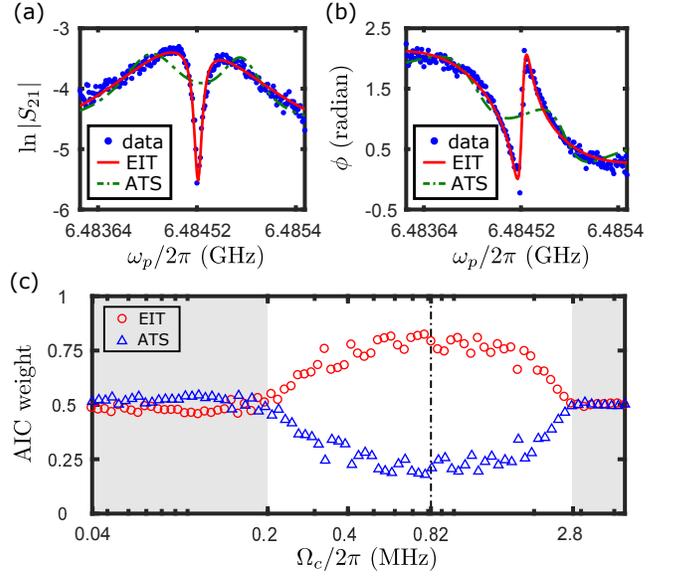}%
 \caption{Fitting the logarithm of transmission magnitude (a) and phase (b) data (blue dots) with the EIT model (red solid curves) and the ATS model (green dash-dotted curves) at $\Omega_c/2\pi=0.82$~MHz. (c) Calculated weights of EIT and ATS by AIC-based testing. Vertical black dash-dotted line is the EIT boundary given by $\Omega_c/2\pi=\gamma_c/2\pi=0.82$~MHz. Shaded regions are where both EIT and ATS model fits do not yield meaningful results. \label{AIC}}
\end{figure}

We fit the probe transmission $S_{21}$ data to both EIT and ATS models to extract the AIC-based testing weights to validate the observations was from EIT \cite{anisimov2011objectively}. For each model, $\ln(|S_{21}|)$ and $\phi$ were fit simultaneously to assure the Kramers-Kronig relations. The transmission data at $\Omega_c/2\pi=0.82$~MHz and its fits of both models are shown in \reffig{AIC}(a)\&(b). Qualitatively, at this control field strength, the data fits significantly better to the EIT model than to the ATS model. Furthermore, the weights of EIT and ATS models for different control field strengths are plotted in \reffig{AIC}(c). For control field strength $\Omega_c/2\pi<0.2$~MHz, both the EIT and ATS weights approach 0.5 due to the presence of noise and the relatively small size of transmission suppressions. In the range of $0.2~\mathrm{MHz}<\Omega_c/2\pi<2.8~\mathrm{MHz}$, the EIT weights are subtantially larger than the ATS weights, indicating strong EIT signatures. The maximum EIT weight happens around $\Omega_c/2\pi=0.82$~MHz, which is in agreement with the theoretical EIT boundary. For control field strength $\Omega_c>2.8$~MHz, the control field excites resonator photons and drives the system out of the nesting regime. Therefore, there is neither EIT nor ATS characters and results in equal weights of 0.5.

 \begin{figure}
 \includegraphics{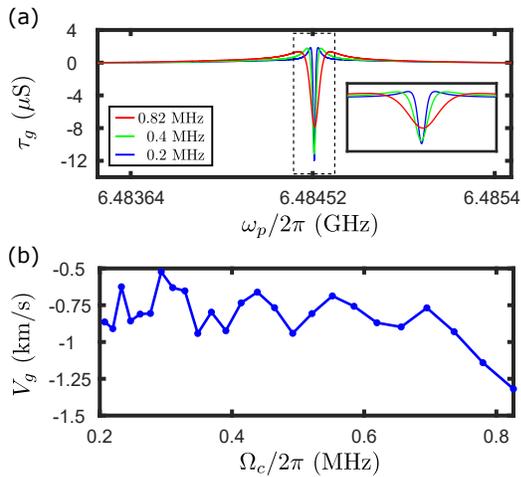}%
 \caption{(a) The time for microwaves to traverse the chip with control field strength, $\Omega_c/2\pi=0.2$~MHz, $0.4$~MHz, and $0.82$~MHz. Inset shows a zoom-in image of (a) around the center frequency. (b) Calculated group velocity in the center of the transmission suppression window as a function of the control field strength.\label{NGV}}
 \end{figure}

We also investigated the backward light phenomenon due to the giant dispersion of EIT \cite{novikov-natphys-2016}. We calculated the time $\tau_g$ for the probe field to traverse the device at different control field strengths by using $\tau_g=-\mathrm{d}\phi/\mathrm{d}\omega_p$, where $\phi$ is obtained from the fittings of the EIT model [\reffig{NGV}(a)]. The group velocity of the probe can then be calculated by $v_g=l/\tau_g$, where $l=10.3$~mm is the distance between the input and output coupler of the device. The largest inferred negative group velocity is $v_g=-0.52\pm0.09$~km/s, further pushing the boundaries of slow light, compared to that reported in reference~\cite{novikov-natphys-2016}.

In conclusion, polariton states in the nesting regime have been generated with a transmon cQED system. The transmission spectra were measured and agree with theoretical predictions. We utilized three levels of nested polariton states to form a $\Lambda$-type transition. A robust EIT signature with all dipole allowed transitions was observed in a superconducting system for the first time. Our results constitute an important step towards scalable quantum network with propagating microwave photons.

\begin{acknowledgments}
X.G. thanks Qi-Chun Liu for discussions. X.G. and Y.X.L. acknowledge the support of the National Basic Research Program of China Grant No. 2014CB921401 and the National Natural Science Foundation of China under Grant No. 91321208. NIST authors acknowledge support of the NIST Quantum Based Metrology Initiative and thank Zachary Dutton for very helpful discussions. This work is property of the US Government and not subject to copyright.
\end{acknowledgments}

\end{document}